\documentclass[aps,prl,twocolumn,superscriptaddress,amsmath,amssymb,floatfix,nofootinbib]{revtex4-2}
\usepackage{amsmath,amssymb}
\usepackage{graphicx}
\usepackage{tikz}

\newcommand{\dgscale}{0.20}
\tikzset{prop/.style={line width=0.5pt}}
\newcommand{\dgvert}{%
  \begin{tikzpicture}[baseline=-0.5ex,scale=\dgscale]
    \draw[prop] (-1,1)--(1,-1); \draw[prop] (-1,-1)--(1,1);
    \fill (0,0) circle (0.13);
  \end{tikzpicture}}
\newcommand{\dgone}{%
  \begin{tikzpicture}[baseline=-0.5ex,scale=\dgscale]
    \draw[prop] (-2,1)--(-1,0)--(-2,-1);
    \draw[prop] (-1,0) arc (180:-180:1);
    \draw[prop] (1,0)--(2,1); \draw[prop] (1,0)--(2,-1);
    \fill (-1,0) circle (0.13); \fill (1,0) circle (0.13);
  \end{tikzpicture}}
\newcommand{\dgtwo}{%
  \begin{tikzpicture}[baseline=-0.5ex,scale=\dgscale]
    \draw[prop] (-3,1)--(-2,0)--(-3,-1);
    \draw[prop] (-2,0) arc (180:-180:1);
    \draw[prop] (0,0) arc (180:-180:1);
    \draw[prop] (2,0)--(3,1); \draw[prop] (2,0)--(3,-1);
    \fill (-2,0) circle (0.13); \fill (0,0) circle (0.13); \fill (2,0) circle (0.13);
  \end{tikzpicture}}
\usepackage{placeins}
\usepackage{xcolor}
\usepackage[colorlinks=true,linkcolor=blue!50!black,citecolor=blue!50!black,urlcolor=blue!50!black]{hyperref}

\newcommand{\nmax}{n_{\max}}
\newcommand{\phib}{\bar\phi}
\newcommand{\VC}{V_{\rm C}}
\newcommand{\lamR}{\lambda_{R}}
\newcommand{\lamE}{\lambda_{R}^{\rm (eff)}}
\newcommand{\SY}{{\rm SY}}
\newcommand{\Om}{\Omega}
\newcommand{\dd}{\mathrm{d}}
\newcommand{\ev}[1]{\langle #1\rangle}

\begin{document}

\title{Lattice Monte Carlo of \texorpdfstring{$\lambda\phi^4$}{lambda phi4} on Euclidean de Sitter}

\author{Angelo Caravano}  
\email{a.caravano@uva.nl}
\affiliation{Gravitation Astroparticle Physics Amsterdam (GRAPPA), University of Amsterdam, Science Park 904, 1098 XH Amsterdam, The Netherlands}

\date{\today}

\begin{abstract}
We report the first Monte Carlo simulation of quantum field theory on Euclidean de Sitter space, $S^4$. Rather than discretizing space, we use a spectral lattice based on spherical harmonics, keeping all symmetries exact at finite cutoff. For a massless $\lambda\phi^4$ theory, the zero-mode distribution reproduces the Starobinsky--Yokoyama equilibrium at weak coupling, and deviations closely follow the one-loop renormalization of $\lambda$. At strong bare coupling, the renormalized quartic coupling saturates near the ceiling of the one-loop renormalization-group flow. This saturation is the de Sitter counterpart of the triviality mechanism that bounds the Higgs mass in flat space. Our results confirm nonperturbatively that the infrared divergences of light scalars in de Sitter are correctly resummed by the stochastic formalism. More broadly, this work opens up de Sitter space to lattice Monte Carlo methods.
\end{abstract}

\maketitle
\emph{Introduction.}---Light scalar fields in de Sitter (dS) space are never weakly interacting in the infrared. For masses below the Hubble rate, $m\ll H$, quantum fluctuations stretched by the expansion exit the Hubble radius $1/H$ and freeze, endlessly feeding the long-wavelength part of the field~\cite{Vilenkin:1982wt,Linde:1982uu,Starobinsky:1982ee}. For a massless field, no dS-invariant Fock vacuum exists~\cite{Allen:1987tz,Allen:1985ux}, and loop corrections around the Bunch--Davies state~\cite{Bunch:1978yq} develop infrared divergences and secular logarithms~\cite{Weinberg:2005vy,Burgess:2010dd,Seery:2010kh}. The standard remedy is stochastic inflation~\cite{Starobinsky:1986fx,Starobinsky:1994bd}: coarse-grained super-Hubble modes obey a Langevin equation whose equilibrium is the Starobinsky--Yokoyama (SY) distribution
\begin{equation}
  P_\SY(\phi)\propto \exp\!\Big[-\frac{8\pi^2}{3H^4}\,V(\phi)\Big].
  \label{eq:SY}
\end{equation}
The stochastic formalism underlies, for example, estimates of Higgs stability during inflation~\cite{Espinosa:2007qp,Markkanen:2018pdo}, of axion and spectator dark matter abundances~\cite{Graham:2018jyp,Markkanen:2018gcw}, of quantum-diffusion tails and primordial black holes~\cite{Pattison:2017mbe,Ezquiaga:2019ftu,Vennin:2024yzl}, and of inflationary correlators~\cite{Vennin:2015hra,Markkanen:2019kpv}.

Confidence in Equation~\eqref{eq:SY} relies on several partial derivations: exact treatment of the Euclidean zero mode on $S^4$~\cite{Rajaraman:2010xd,Beneke:2012kn,LopezNacir:2016gzi}, large-$N$ and nonperturbative renormalization-group resummations~\cite{Serreau:2011fu,Serreau:2013psa,Guilleux:2015pma,LopezNacir:2019ord}, and perturbative and effective-field-theory derivations of the infrared dynamics~\cite{Garbrecht:2014dca,Tokuda:2017fdh,Miao:2026cjj,Gorbenko:2019rza,Mirbabayi:2019qtx,Pinol:2020cdp,Cohen:2020php,Cohen:2021fzf,Cable:2022uwd,Beneke:2026rtf}. All treat the longest modes nonperturbatively but expand or truncate the rest, and the status of infrared secular terms is still disputed~\cite{Huenupi:2024ksc,Palma:2025oux}. What is missing is what lattice QCD provides for hadrons~\cite{Wilson:1974sk,Durr:2008zz,Fodor:2012gf}: a first-principles computation in which \emph{all} modes are integrated with the full quantum-field-theory weight. On Euclidean dS---the four-sphere $S^4$---that weight is real and positive, so it can be sampled with Monte Carlo methods. The obstacle is geometric, since $S^4$ admits no refinable regular discretization~\cite{Coxeter:1973,Brower:2016moq,VicenteGarcia-Consuegra:2025two} and a simplicial lattice breaks the dS isometries at finite spacing.

In this work we remove that obstacle with a spectral regulator that keeps the dS isometry group $\mathrm{SO}(5)$ exact at finite cutoff, and use it to perform what is, to our knowledge, the first nonperturbative Monte Carlo simulation of quantum field theory on Euclidean dS. Monte Carlo studies of quantum fields in curved spacetime have so far addressed hypercubic lattices with a time-dependent scale factor~\cite{Yamamoto:2014vda} or lower-dimensional geometries: lattice discretizations of $S^2$ and $\mathbb{R}\times S^2$~\cite{Brower:2016moq,Brower:2018szu,Brower:2012vg,Brower:2020jqj}, tessellations of hyperbolic space and AdS$_2$~\cite{Asaduzzaman:2020hjl,Brower:2019kyh}, and noncommutative fuzzy-sphere regularizations~\cite{Panero:2006bx,GarciaFlores:2009hf}. 

For a massless field, we find that (i) the SY distribution, including its non-Gaussian shape, is recovered at the few-per-mille level at weak coupling, without ever invoking a Langevin equation; (ii) deviations from it are fixed, with no free parameter, by the one-loop renormalization of the quartic coupling, following the flat-space $\beta$ function; and (iii) at strong bare coupling, the renormalized quartic coupling saturates at a finite value, while the zero-mode distribution stays close to SY. This is the same screening mechanism behind the lattice upper bound on the Higgs mass in flat space~\cite{Luscher:1987ay,Kuti:1987nr}. Triviality, usually a statement about the ultraviolet, thus protects the infrared predictions of stochastic inflation. 

\begin{figure*}[t]
  \centering
  \includegraphics[width=\textwidth]{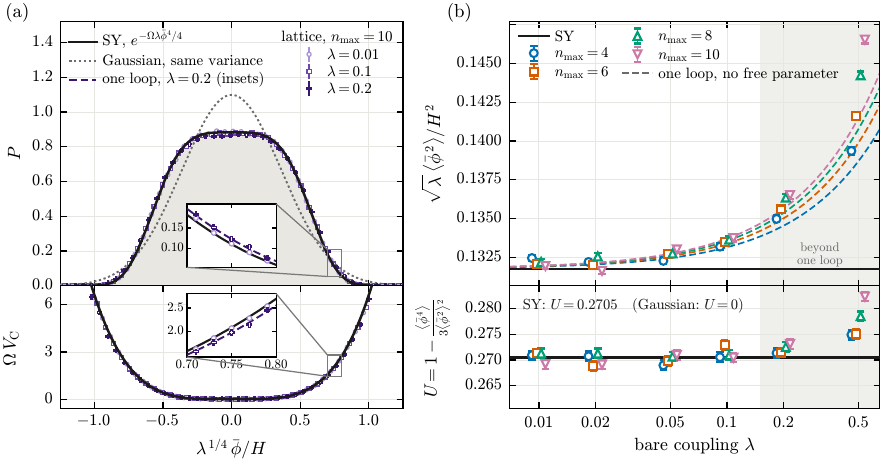}
  \caption{Starobinsky--Yokoyama (SY) at weak coupling, with the bare mass set to its one-loop counterterm, so that the renormalized mass $m_{R}^2$ vanishes at one loop; (a)~Top: zero-mode distribution at $\nmax=10$ for $\lambda=0.01$, $0.1$ and $0.2$, against SY (solid). Bottom: the same data as $\Om\VC=-\ln P+\text{const}$, against the tree-level quartic. The zoom-in insets show the one-loop departure from tree level at $\lambda=0.2$ (dashed). (b)~Top: $\sqrt\lambda\ev{\phib^2}$ at four cutoffs against the SY prediction $0.13176\,H^2$ (solid). Dashed lines are the one-loop predictions of Eq.~\eqref{eq:prediction}. Bottom: the Binder cumulant $U=1-\ev{\phib^4}/3\ev{\phib^2}^2$ of the same data against $U_\SY=0.27052$ (solid). Shaded: region where the one-loop description ceases to be accurate.}
  \label{fig:sy}
\end{figure*}

\emph{Regulated theory on $S^4$.}---We study a real scalar field with quartic interaction on a fixed dS background of Hubble rate $H$. Under the Wick rotation $t\to-\mathrm{i}\tau$, dS space becomes the four-sphere of radius $1/H$ and volume $\Om=8\pi^2/3H^4$~\cite{Gibbons:1977mu}. This analytic continuation selects one particular state of the theory, the Euclidean vacuum~\cite{Hartle:1983ai,Marolf:2010nz,Hollands:2010pr}, which reduces to the Bunch--Davies state~\cite{Bunch:1978yq} for free massive fields. The action is
\begin{equation}
  S=\int_{S^4}\!\dd^4x\sqrt{g}\,\Big[\tfrac12(\nabla\phi)^2
     +\tfrac12 m^2\phi^2+\tfrac{\lambda}{4}\phi^4\Big],
  \label{eq:action}
\end{equation}
where a curvature coupling $\xi R\phi^2$ is absorbed into $m^2$. We expand in real hyperspherical harmonics~\cite{Higuchi:1986wu},
\begin{equation}
  \phi(x)=\sum_{n=0}^{\nmax}\sum_{I=1}^{d_n} c_{nI}\,Y_{nI}(x),
  \label{eq:modes}
\end{equation}
where $-\nabla^2Y_{nI}=\lambda_nY_{nI}$ with $\lambda_n=n(n+3)H^2$, and each level $n$, of dimension $d_n=(2n+3)(n+1)(n+2)/6$, is a complete irreducible representation of $\mathrm{SO}(5)$. The free action is diagonal, $S_0=\frac12\sum(\lambda_n+m^2)c_{nI}^2$. The interaction is computed pseudo-spectrally, by evaluating $\phi^4$ on a real-space grid large enough that the interaction term is computed exactly, without aliasing (Appendix~\ref{app:reg}). The regulated theory is then \emph{exactly} $\mathrm{SO}(5)$ invariant, which we verify directly by rotating the grid by random $\mathrm{SO}(5)$ elements and checking that the action is invariant (Appendix~\ref{app:reg}). The cutoff of the theory is $\Lambda=H\sqrt{\nmax(\nmax+3)}\simeq\nmax H$, and the number of modes is $N=(\nmax+1)(\nmax+2)^2(\nmax+3)/12$, e.g. $1716$ for $\nmax=10$.

\emph{Zero mode and the constraint effective potential.}---The bridge between the Euclidean theory and stochastic inflation is the distribution of the four-volume average, or homogeneous mode, $\phib\equiv\Om^{-1}\!\int\!\sqrt{g}\,\phi$,
\begin{equation}
  P(\phib)\equiv e^{-\Om\,\VC(\phib)}\propto
  \int\!\mathcal{D}\phi\;\delta\Big(\phib-\frac{1}{\Om}\!\int\!\sqrt g\,\phi\Big)\,e^{-S[\phi]},
  \label{eq:VC}
\end{equation}
which defines the constraint effective potential $\VC$~\cite{Fukuda:1974ey,ORaifeartaigh:1986axd}. Recent work provided evidence that $\VC$, rather than the standard effective potential~\cite{Coleman:1973jx}, is the quantity entering the stochastic theory~\cite{Camargo-Molina:2022paw,VicenteGarcia-Consuegra:2025two}. The two coincide at tree level, but the loop expansion of the standard effective potential is infrared singular for light fields~\cite{Serreau:2011fu,Rajaraman:2010xd,Gorbenko:2019rza}, while $\VC$ is finite because the constant mode is held fixed rather than integrated over. At one loop, Eq.~\eqref{eq:VC} reduces to a Gaussian integral over the nonzero modes, which gives
\begin{equation}
  \VC(\phib)=V(\phib)+\frac{1}{2\Om}\sum_{n=1}^{\nmax}d_n
  \ln\!\Big[1+\frac{V''(\phib)}{\lambda_n}\Big]+\dots,
  \label{eq:VC1loop}
\end{equation}
the spectral-cutoff form of the result of Ref.~\cite{VicenteGarcia-Consuegra:2025two}, with the sum starting at $n=1$. In the simulations $P(\phib)$ is measured directly as the histogram of $\phib$ over configurations of the full theory. Therefore, any departure from Eq.~\eqref{eq:SY} is a physical effect of the $n\ge1$ modes, and can be compared with the one-loop prediction of Eq.~\eqref{eq:VC1loop}.

\emph{One-loop predictions.}---Expanding Eq.~\eqref{eq:VC1loop} in powers of $3\lambda\phib^2/\lambda_n$ is a standard one-loop computation, which, because $\phib^2\sim H^2/\sqrt{\lambda}$ from Eq.~\eqref{eq:SY}, organizes contributions in powers of $\sqrt\lambda$~\cite{Rajaraman:2010xd,Beneke:2012kn}. With our regulator, every coefficient in this expansion can be evaluated explicitly, so that loop predictions can be tested by the Monte Carlo without free parameters. Three terms matter.

(i)~\emph{Mass.} The term quadratic in $\phib$ shifts the mass, $m_R^2=m^2+3\lambda S_1/\Om$, where $S_1=\sum_{n=1}^{\nmax}d_n/\lambda_n\propto\nmax^2$, so that $S_1/\Om$ is the coincident propagator \emph{without} the zero mode~\cite{Rajaraman:2010xd,Beneke:2012kn}. In this work we aim to study this system in the massless limit $m_R=0$. At weak coupling we set the bare mass to the analytic one-loop counterterm $m^2=-3\lambda S_1/\Om$, which requires no measurement and gives $m_R^2=O(\lambda^2)$. At strong coupling this estimate is no longer sufficiently accurate, so instead we will impose $m_R^2\equiv\VC''(0)=0$ nonperturbatively, measuring the curvature of $\VC$ directly.

(ii)~\emph{Coupling.} The renormalized coupling $\lamR$ is defined as the coefficient of $\phib^4/4$ in $\VC$, which at one loop gives
\begin{align}
  \lamR&=\lambda\big[1-u\lambda\big],\qquad u\equiv\frac{9S_2(\nmax)}{\Om},\nonumber\\
  S_2&=\sum_{n=1}^{\nmax}\frac{d_n}{\lambda_n^2}\simeq\frac{1}{3H^4}\ln\nmax,
  \label{eq:screen}
\end{align}
and its cutoff dependence, carried by the screening coefficient $u$, is the running of $\lambda$. At large cutoff $u\simeq(9/8\pi^2)\ln(\Lambda/H)$, matching the flat-space $\beta$ function $\beta_\lambda=9\lambda^2/8\pi^2$.

(iii)~\emph{Beyond quartic order.} The next term adds a sextic interaction $\frac92\lambda^3S_3\,\phib^6$ to $\Om\VC$, with $S_3(\nmax)=\sum_{n=1}^{\nmax}d_n/\lambda_n^3\simeq0.10\,H^{-6}$ nearly independent of the cutoff, since the sum converges; it gives corrections of relative order $\lambda^{3/2}$.

The massless SY distribution has
\begin{align}
  \ev{\phib^2}_\SY&=\sqrt{\frac{3}{2\pi^2\lambda}}\,
  \frac{\Gamma(\frac34)}{\Gamma(\frac14)}\,H^2\simeq\frac{0.13176\,H^2}{\sqrt\lambda},
  \nonumber\\
  U_\SY&=1-\frac{\Gamma(\frac14)^2}{12\,\Gamma(\frac34)^2}\simeq0.2705,
  \label{eq:SYnumbers}
\end{align}
where $U=1-\ev{\phib^4}/3\ev{\phib^2}^2$ is the Binder cumulant ($U=0$ for a Gaussian). Evaluating SY at the one-loop-renormalized parameters, $m_R^2=0+O(\lambda^2)$ and $\lamR$ of Eq.~\eqref{eq:screen}, and comparing with SY at the bare coupling gives the parameter-free prediction
\begin{align}
  \frac{\ev{\phib^2}}{\ev{\phib^2}_\SY}-1&=\frac{u}{2}\,\lambda
  +O(\lambda^{3/2}),\nonumber\\
  U-U_\SY&=O(\lambda^{3/2}).
  \label{eq:prediction}
\end{align}
Through relative order $O(\lambda)$, loop corrections thus renormalize the parameters of SY without changing the shape of $P(\phib)$; the first shape corrections enter at relative order $O(\lambda^{3/2})$. In the leading SY zero-mode action at $m_R^2=0$, the quartic is the only term, so $\ev{\phib^2}\propto\lambda^{-1/2}$ is nonperturbative and inaccessible to any finite order of perturbation theory around the Gaussian~\cite{Rajaraman:2010xd,Beneke:2012kn}.\footnote{Weak and strong coupling refer to the size of the bare coupling $\lambda$.}

\emph{Monte Carlo.}---The regulated path integral is an ordinary integral over the $N$ mode coefficients $c_{nI}$ with positive weight $e^{-S}$, which we sample with hybrid Monte Carlo~\cite{Duane:1987de}, preconditioned with the free propagator (Fourier acceleration~\cite{Batrouni:1985jn}) and with trajectory lengths randomized by $\pm30\%$ to avoid resonances~\cite{Mackenzie:1989us}. The zero mode, the slowest direction, is additionally resampled after each trajectory from its conditional distribution at fixed nonzero modes. The algorithm and its validation are described in Appendix~\ref{app:mc}; statistical errors are estimated with autocorrelation and blocked-resampling methods~\cite{Wolff:2003sm}. The code developed for this work, \href{https://github.com/caravangelo/lattice-dS}{\texttt{lattice-dS}}, is publicly available, and an animation of the sampling can be found \href{https://caravangelo.github.io/lattice-dS/}{here}.

We use two sets of simulations, which differ in how the massless point $m_R^2=0$ is reached. In Set~A ($\lambda=0.01$--$0.5$, $\nmax=4$--$10$) the bare mass is fixed in advance to the one-loop counterterm $m^2=-3\lambda S_1/\Om$; Results~1 and~2 use this set (Appendix~\ref{app:data}). In Set~B ($\lambda=0.2$--$60$, $\nmax=4$--$10$) the one-loop estimate is no longer accurate and the massless point is determined nonperturbatively, as described next; Result~3 uses this set (Appendix~\ref{app:setb}).

\emph{Nonperturbative mass and coupling.}---At strong coupling we determine $m_R^2$ and $\lamR$ directly from the simulations, borrowing the mean-force technique of constrained molecular dynamics~\cite{Kirkwood:1935pmf,Carter:1989crc,Sprik:1998fcm}, also used for lattice gauge-Higgs models~\cite{Gunther:2019feg}. We use \emph{constrained} simulations, where the zero mode is held fixed at a chosen $\phib$ and only the other modes are sampled. Such a simulation gives direct access to $\VC'(\phib)$ through the exact identity
\begin{equation}
  \Om\,\VC'(\phib)=\Big\langle\frac{\partial S}{\partial\phib}\Big\rangle_{\phib},
  \label{eq:force}
\end{equation}
where $\ev{\cdot}_{\phib}$ is the average in the constrained ensemble (Appendix~\ref{app:force}). Thus $\VC'$ is the mean force on the zero mode.

Constrained simulations at $\phib=0$ and at small $\phib$ give the curvature $m_R^2=\VC''(0)$ at the simulated bare mass. Since the action is linear in $m^2$, configurations sampled at one bare mass $m^2$ can be reweighted to another bare mass $m^2+\delta m^2$~\cite{Ferrenberg:1988yz,Hasenfratz:2008fg}. This gives the function $m_R^2(m^2)$ without running new simulations; the massless point is found numerically as the zero of this function, at which all observables are evaluated (Appendix~\ref{app:massless}).

At strong coupling $\VC$ is no longer a pure quartic, and we define the renormalized coupling from the mean force at the typical zero-mode value:
\begin{equation}
  \lamE\equiv\frac{\VC'(\sigma)}{\sigma^3},\qquad \sigma^2=\ev{\phib^2},
  \label{eq:lameff}
\end{equation}
where $\sigma$ is obtained from unconstrained simulations reweighted to the massless point. For a pure quartic $\VC$, $\lamE$ coincides with the coefficient of $\phib^4/4$ that defines $\lamR$; in general, the two differ only through higher-order terms, which become important at strong coupling. For more details about this procedure, see Appendix~\ref{app:setb}.

\emph{Result 1: Starobinsky--Yokoyama is recovered.}---Figure~\ref{fig:sy}(a) shows the zero-mode distribution from the weak-coupling set. The agreement with SY is extremely accurate: in the upper panel the lattice histograms for $\lambda=0.01$, $0.1$ and $0.2$ follow the SY distribution, flat-topped and with sharper tails than a Gaussian of the same width. Plotted as $-\ln P$ (lower panel), the same data are a direct Monte Carlo measurement of the constraint effective potential. Figure~\ref{fig:sy}(b) makes this quantitative through the two lowest moments. At the weakest coupling, $\lambda=0.01$, the variance agrees with SY to $0.5\%$ at every cutoff. The Binder cumulant, which measures the shape, agrees with $U_\SY=0.2705$ to within $1\%$ up to $\lambda=0.2$, where a Gaussian would give $U=0$, and deviates by up to $4.4\%$ at $\lambda=0.5$.

\begin{figure}[t]
  \centering
  \includegraphics[width=\columnwidth]{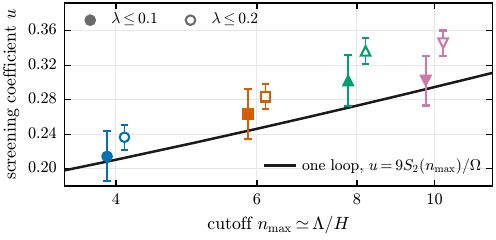}
  \caption{The screening coefficient $u$, measured from the residual from SY, $\ev{\phib^2}/\ev{\phib^2}_\SY-1=u\lambda/2$, against the cutoff. The line is its one-loop value $u=9S_2/\Om$, Eq.~\eqref{eq:screen}, with no free parameter. Filled symbols are fitted on $\lambda\le0.1$, open symbols on $\lambda\le0.2$, where the higher-order correction has set in.}
  \label{fig:run}
\end{figure}

\emph{Result 2: the residual is the running coupling.}---The deviation of the variance from SY, shown in Fig.~\ref{fig:sy}(b), is positive and grows with both the coupling and the cutoff. At weak coupling, Eq.~\eqref{eq:prediction} predicts that it grows as $u\lambda/2$, so a linear fit in $\lambda$ at each cutoff measures the screening coefficient $u$. On $\lambda\le0.1$ the measured values agree with the one-loop $u$ at every cutoff, as shown in Fig.~\ref{fig:run}. A common fit of $u=\kappa\,9S_2/\Om$ on $\lambda\le0.1$ gives $\kappa=1.06\pm0.06$ against the one-loop value $1$. Since $S_2\simeq\frac{1}{3H^4}\ln\nmax$, this is the logarithmic running of the quartic coupling, obtained from the lattice without free parameters. 

Widening the window to $\lambda\le0.2$ (open symbols in Fig.~\ref{fig:run}) makes the measured values of $u$ exceed the prediction. The excess is caused by higher-order corrections, which make the analytic mass tuning no longer accurate enough. Beyond $\lambda\simeq0.1$ the bare mass therefore has to be tuned nonperturbatively, which we do in the strong-coupling simulations presented next.

\begin{figure}[t]
  \centering
  \includegraphics[width=\columnwidth]{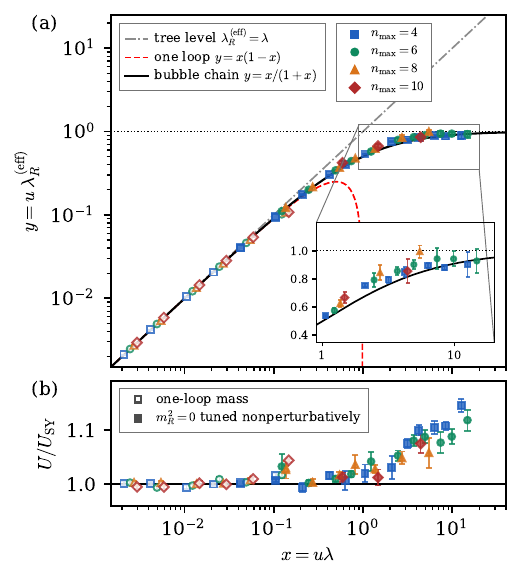}
  \caption{Renormalized coupling and Binder cumulant over four decades of bare coupling; open symbols use the one-loop mass, filled symbols the renormalized mass set to zero nonperturbatively. (a)~The effective quartic coupling of the zero mode, Eq.~\eqref{eq:lameff}, in the variables $x=u\lambda$, $y=u\lamE$, against tree level ($y=x$), the one-loop truncation $y=x(1-x)$ and the resummed chain $y=x/(1+x)$ of Eq.~\eqref{eq:chain}. (b)~The Binder cumulant relative to $U_\SY$ at the same points.}
  \label{fig:triv}
\end{figure}

\emph{Result 3: screening follows the one-loop flow.}---Equation~\eqref{eq:screen} is the first-order solution of the one-loop renormalization-group flow $\dd\lamR/\dd u=-\lamR^2$~\cite{Gell-Mann:1954yli,Wilson:1973jj}. Integrating that flow rather than expanding it, which resums the chain of one-loop bubbles, gives
\begin{equation}
  \lamR=\dgvert\,+\,\dgone\,+\,\dgtwo\,+\,\cdots=\frac{\lambda}{1+u\lambda},
  \label{eq:chain}
\end{equation}
or $1/\lamR=1/\lambda+u$, so that $1/\lamR$ is linear in $\ln\Lambda$ and $\lamR$ approaches $1/u$ as $\lambda\to\infty$.

Equation~\eqref{eq:chain} keeps one class of diagrams. That is justified when a large logarithm makes the leading logarithms dominate~\cite{Coleman:1973jx}, or when a large number of fields $N$ makes the bubbles dominate~\cite{Coleman:1974jh}; here $\ln(\Lambda/H)\simeq1.7$--$2.4$ and $N=1$. This means that, already at $x\equiv u\lambda\sim1$, the neglected two-loop diagrams are of the same order as the second bubble, and nothing guarantees that Eq.~\eqref{eq:chain} holds. Whether the coupling nevertheless saturates is the dS form of the question behind the lattice upper bound on the Higgs mass~\cite{Dashen:1983ts,Luscher:1987ay,Luscher:1987ek,Luscher:1988uq,Kuti:1987nr,Callaway:1988ya,Gerhold:2010bh}. Since $\phi^4_4$ is trivial, i.e.\ a free theory for $\Lambda\rightarrow\infty$, the renormalized coupling is bounded at finite cutoff~\cite{Aizenman:2019yuo}. In flat space the largest coupling at a given cutoff is obtained at $\lambda=\infty$, where the field sits at a minimum of the potential at every lattice site and the theory is an Ising model (or, for the four-component Higgs field, a non-linear sigma model)~\cite{Luscher:1987ay,Luscher:1987ek,Luscher:1988uq}. In our spectral regularization this endpoint is not directly accessible, since a truncated field cannot obey a pointwise constraint. The large-$\lambda$ limit must instead be approached from below, pushing $\lambda$ as far as the technique allows.

Figure~\ref{fig:triv}(a) shows the value of $\lamE$ extracted from our simulations, plotted as $y=u\lamE$ against $x=u\lambda$. At weak coupling $\lamE$ coincides with the bare coupling and then follows the one-loop screening of Eq.~\eqref{eq:screen}, to within $3\%$ for $\lambda\le0.2$. Beyond $x\simeq0.3$ the one-loop truncation fails, while the data follow the bubble chain and saturate. Above $x\simeq4$, the points at all four cutoffs lie on a common plateau near the ceiling $1/u$. A constant fit to the ten points with $x>4$ gives $y_\infty=0.89\pm0.01$ with $\chi^2/{\rm dof}=9.6/9$, stable within errors for cuts between $x=3.5$ and $6$. Over the whole range, Eq.~\eqref{eq:chain} describes the data to within about $20\%$, up to bare couplings two orders of magnitude beyond the one-loop regime.

The Binder cumulant, measured from $P(\phib)$ at the massless point and shown in Fig.~\ref{fig:triv}(b), stays within $12\%$ of $U_\SY$ up to $x\simeq10$, and within $15\%$ up to $x\simeq15$, at every cutoff. This shows that the distribution approximately keeps the SY shape in the strong-coupling regime. As a consequence, the saturation also bounds the mass the field generates for itself. In the stochastic formalism, the zero mode relaxes to the SY equilibrium at the rate $\Gamma=0.0889\sqrt{\lamR}\,H$~\cite{Starobinsky:1994bd}; defining $m_{\rm dyn}^2\equiv3H\Gamma$ and inserting the measured coupling, we find $m_{\rm dyn}<0.8\,H$ at all cutoffs studied, no matter how large the bare coupling. This bound relies on the stochastic formalism: $\Gamma$ is a dynamical quantity whose direct computation is a Lorentzian question beyond our Euclidean treatment, which we leave for future work.

\emph{Conclusions.}---We have developed a lattice Monte Carlo formulation of quantum field theory on Euclidean dS space and applied it to massless $\lambda\phi^4$ theory. With all modes retained nonperturbatively, the zero-mode distribution reproduces Starobinsky--Yokoyama at the few-per-mille level at weak coupling, and as the coupling grows the deviations from it follow the one-loop prediction. At large bare coupling, up to $\lambda=60$, the SY shape is approximately preserved, with $\lambda$ replaced by the renormalized quartic coupling that we measure nonperturbatively. The coupling saturates, as predicted by the one-loop renormalization-group flow, well beyond the regime in which the flow is justified. This screening is the dS analogue of the mechanism behind the triviality bound on the Higgs mass. The Starobinsky--Yokoyama distribution therefore remains a good approximation even in the strong-coupling regime. 

These results are also relevant to the interpretation of infrared secular terms, whose status remains debated~\cite{Huenupi:2024ksc,Palma:2025oux}. Our simulations demonstrate nonperturbatively that the SY distribution is the equilibrium of the full quantum theory, indicating that these terms are correctly resummed by stochastic inflation. It should be stressed, however, that Euclidean $S^4$ describes the equilibrium state of the theory. Relaxation towards it is a Lorentzian question that we do not treat, and the background is fixed, so neither gravitational backreaction~\cite{Moreau:2018lmz} nor the quantum stability of dS~\cite{Dvali:2017eba} is addressed.

Several questions remain open. The most direct is the continuum limit. At the cutoffs reached here, the logarithm that drives the screening is $\ln(\Lambda/H)\simeq1.7$--$2.4$. As the cutoff grows, the ceiling $1/u$ should decrease as $1/\ln(\Lambda/H)$, making the triviality of $\phi^4_4$ visible directly on dS as $\lamR\to0$, as in flat space~\cite{Luscher:1987ay,Luscher:1987ek}. Since the number of modes grows as $\nmax^4$, reaching the continuum limit requires more efficient sampling, perhaps using normalizing flows~\cite{Albergo:2019eim,Cheng:2025khv}. Another question is the $\lambda=\infty$ endpoint, which a spectral cutoff makes difficult to study; a regulator that preserves the pointwise strong-coupling constraint could turn the observed saturation into a bound read off directly at $\lambda=\infty$, as in flat space.

Beyond these results, this work shows that lattice field theory on Euclidean dS space is possible, putting a range of nonperturbative dS questions within computational reach. For instance, tunneling can be addressed: at negative $m_R^2$ the barrier that controls it is the constraint effective potential~\cite{Camargo-Molina:2022paw}, which the simulations measure directly. Other interesting targets are $O(N)$ models~\cite{Serreau:2013psa,LopezNacir:2019ord}, and a confrontation with the $\sqrt\lambda$ expansion~\cite{Beneke:2012kn} and soft de Sitter effective theory~\cite{Cohen:2021fzf,Beneke:2026rtf}. Moreover, since $S^4$ harmonics contain those of $S^3$, the same regulator can be applied to the spatial slices of Lorentzian dS in global coordinates. We leave these exciting directions for future work.

\section*{Data and code availability}
The Monte Carlo time series of every simulation, the tables underlying all figures, and a script that reproduces Figs.~1--3 from these tables are available on Zenodo at \href{https://doi.org/10.5281/zenodo.23242257}{doi:10.5281/zenodo.23242257}. The simulation and analysis code, together with the scripts that recompute the tables from the time series and produce the figures of the appendices, is available at \url{https://github.com/caravangelo/lattice-dS}.

\section*{Acknowledgments}
\begin{acknowledgments}
I thank Daniel Baumann, Priyesh Chakraborty, Ben Freivogel, Arttu Rajantie, S\'ebastien Renaux-Petel, Jan Pieter van der Schaar and Erik Verlinde for interesting discussions. This project has received funding from the European Union's Horizon Europe research and innovation programme under the Marie Sk{\l}odowska-Curie grant agreement No.~101202657. 
\end{acknowledgments}

\bibliography{refs}

\onecolumngrid
\appendix
\setcounter{secnumdepth}{3}

\section{Spectral regulator}
\label{app:reg}

The regulator retains complete multiplets of harmonics and evaluates the interaction in real space without aliasing, keeping $\mathrm{SO}(5)$ exact at finite cutoff. We set $H=1$, so $\Om=8\pi^2/3$.

\subsection{Regulated action}

We start from the action of Eq.~\eqref{eq:action} on the unit four-sphere, in the coordinates
\begin{equation}
 \dd s^2=\dd\chi^2+\sin^2\!\chi\,\big[\dd\psi^2+\sin^2\!\psi\,(\dd\theta^2+\sin^2\!\theta\,\dd\varphi^2)\big],
 \qquad \sqrt g=\sin^3\!\chi\,\sin^2\!\psi\,\sin\theta ,
\end{equation}
with $\chi,\psi,\theta\in[0,\pi]$ and $\varphi\in[0,2\pi)$. The real hyperspherical harmonics are~\cite{Higuchi:1986wu}
\begin{equation}
 Y_{nLlm}=N_{nL}\sin^L\!\chi\,C^{(L+3/2)}_{n-L}(\cos\chi)\;
 N_{Ll}\sin^l\!\psi\,C^{(l+1)}_{L-l}(\cos\psi)\;Y_{lm}(\theta,\varphi),
 \label{eq:app-harmonics}
\end{equation}
with integer labels $n\ge L\ge l\ge|m|$, where $C^{(\alpha)}_k$ are Gegenbauer polynomials~\cite{Olver:2010ouy} and $Y_{lm}$ are the real orthonormal spherical harmonics on $S^2$. The normalization constants $N_{nL}$ and $N_{Ll}$, given by
\begin{equation}
 N_{nL}^{-2}=h^{(L+3/2)}_{n-L},\quad
 N_{Ll}^{-2}=h^{(l+1)}_{L-l},\quad
 h^{(\alpha)}_k=\int_{-1}^{1}\!\dd\eta\,(1-\eta^2)^{\alpha-\frac12}\big[C^{(\alpha)}_k(\eta)\big]^2
 =\frac{\pi\,2^{1-2\alpha}\,\Gamma(k+2\alpha)}{k!\,(k+\alpha)\,\Gamma(\alpha)^2},
 \label{eq:app-norm}
\end{equation}
make the basis orthonormal, $\int\!\sqrt g\,Y_{nI}Y_{n'I'}=\delta_{nn'}\delta_{II'}$, where $I=(L,l,m)$. Each harmonic satisfies $-\nabla^2Y_{nI}=\lambda_nY_{nI}$ with $\lambda_n=n(n+3)$, and at fixed $n$ the index $I$ takes $\sum_{L=0}^{n}\sum_{l=0}^{L}(2l+1)=d_n$ values, which span an irreducible representation of $\mathrm{SO}(5)$. The sum over a level obeys the addition theorem~\cite{Frye:2012jj}
\begin{equation}
 \sum_IY_{nI}(x)Y_{nI}(y)=\frac{d_n}{\Om}
 \frac{C_n^{(3/2)}(x\cdot y)}{C_n^{(3/2)}(1)},
 \label{eq:app-addition}
\end{equation}
where $x\cdot y$ is the inner product of the two points embedded in $\mathbb R^5$. Numerically, Eq.~\eqref{eq:app-addition} holds to $4\times10^{-15}$, which verifies the normalization.

The regulator truncates the expansion of Eq.~\eqref{eq:modes} at $n\le\nmax$. Since only complete levels are kept, the retained mode space is $\mathrm{SO}(5)$ invariant. By orthonormality, the kinetic and mass terms are diagonal in the mode coefficients, and the regulated action is
\begin{equation}
 S=\frac12\sum_{n=0}^{\nmax}\sum_I(\lambda_n+m^2)c_{nI}^2+\frac\lambda4\int_{S^4}\!\sqrt g\,\phi^4 .
 \label{eq:app-action}
\end{equation}
The quartic term couples all retained modes and is the only part of the action that requires numerical integration.

\subsection{Exact quadrature of the interaction}

The dynamical variables are the mode coefficients $c_{nI}$, and the quartic term is evaluated pseudo-spectrally. The field is reconstructed in real space, $\phi(x_a)=\sum_{nI}c_{nI}Y_{nI}(x_a)$, at the nodes $x_a$ of a quadrature grid on $S^4$; there $\phi^4$ is computed pointwise and integrated with weights $w_a$, defined below. The nodes are not lattice sites: they carry no degrees of freedom of their own, and the field on them is fixed by the coefficients $c_{nI}$.

For a function $f(\chi,\psi,\theta,\varphi)$ on $S^4$, the integral is written as nested one-dimensional integrals over the four angles, and each is replaced by its own one-dimensional quadrature,
\begin{equation}
 \int_{S^4}\!\sqrt g\,f
 =\int_0^\pi\!\dd\chi\,\sin^3\!\chi\int_0^\pi\!\dd\psi\,\sin^2\!\psi\int_0^\pi\!\dd\theta\,\sin\theta\int_0^{2\pi}\!\dd\varphi\,f
 \;\to\;\sum_{i,j,k,l}w^\chi_i\,w^\psi_j\,w^\theta_k\,w^\varphi_l\,f(\chi_i,\psi_j,\theta_k,\varphi_l).
 \label{eq:app-product}
\end{equation}
In terms of $\eta_\chi=\cos\chi$, $\eta_\psi=\cos\psi$ and $\eta_\theta=\cos\theta$, this integral becomes
\begin{equation}
 \int_{S^4}\!\sqrt g\,f=\int_{-1}^{1}\!\dd\eta_\chi\,(1-\eta_\chi^2)\int_{-1}^{1}\!\dd\eta_\psi\,(1-\eta_\psi^2)^{1/2}\int_{-1}^{1}\!\dd\eta_\theta\int_0^{2\pi}\!\dd\varphi\,f .
 \label{eq:app-weights}
\end{equation}
Each polar integral thus has the form $\int_{-1}^{1}\dd\eta\,\rho(\eta)\,g(\eta)$, with $\rho(\eta)=1-\eta^2$, $(1-\eta^2)^{1/2}$ and $1$ for $\eta_\chi$, $\eta_\psi$ and $\eta_\theta$, respectively, and $g$ the rest of the integrand. We replace it by $\sum_{i=1}^{N_p}w_i\,g(\eta_i)$, using for each $\rho$ the $N_p$ nodes $\eta_i$ and weights $w_i$ of Gaussian quadrature, which make the sum exact for every polynomial $g$ of degree up to $2N_p-1$~\cite{Olver:2010ouy}. The polar nodes are then $\chi_i=\arccos \eta_{\chi,i}$, where $\eta_{\chi,i}$ and $w^\chi_i$ are the nodes and weights for $\rho=1-\eta^2$, and similarly $\psi_j$, $w^\psi_j$ and $\theta_k$, $w^\theta_k$. The azimuthal integral is replaced by a sum over $N_\varphi$ equally spaced nodes $\varphi_l=2\pi l/N_\varphi$ with $w^\varphi_l=2\pi/N_\varphi$, which is exact for trigonometric polynomials of degree below $N_\varphi$. The grid points $x_a$ are the $N_p^3N_\varphi$ combinations $(\chi_i,\psi_j,\theta_k,\varphi_l)$, with weights $w_a=w^\chi_iw^\psi_jw^\theta_kw^\varphi_l$. The points are not uniformly distributed on the sphere and accumulate where the sine factors of the measure vanish; the weights compensate for this exactly.

A hyperspherical harmonic $Y_{nI}$ of level $n$ is the restriction to $S^4$ of a homogeneous polynomial of degree $n$ in $\mathbb R^5$~\cite{Frye:2012jj}. A field truncated at $\nmax$ is therefore a polynomial of degree at most $\nmax$, and both $\phi^4$ and $Y_{nI}\phi^3$, for any retained mode, have degree at most $4\nmax$. We say that a grid has degree $D$ if it integrates exactly all polynomials of degree up to $D$. A grid with $D\ge4\nmax$ gives
\begin{equation}
 S_{\rm int}=\frac\lambda4\sum_a w_a\phi(x_a)^4,
 \qquad
 \frac{\partial S_{\rm int}}{\partial c_{nI}}
 =\lambda\sum_a w_aY_{nI}(x_a)\phi(x_a)^3 ,
 \label{eq:app-dealias}
\end{equation}
which are then the continuum interaction and its derivative within the retained mode space, free of aliasing and exact up to floating-point error.

We use grids of degree $D=4\nmax$, the smallest for which the interaction is evaluated without aliasing; this fixes the numbers of nodes. A polynomial of degree at most $D$ in the coordinates of $\mathbb R^5$, written in the angles, reduces after the $\varphi$ sum to polynomials of degree at most $D$ in $\eta_\chi$, $\eta_\psi$ and $\eta_\theta$ against the weights of Eq.~\eqref{eq:app-weights}, since terms with odd powers of the sines cancel. The grid therefore has degree $D$ if $2N_p-1\ge D$ and $N_\varphi>D$, which gives $N_p=2\nmax+1$ and $N_\varphi=4\nmax+1$. At $\nmax=10$ this is $21^3\times41=379\,701$ points for $1716$ modes.

\begin{figure}
  \centering
  \includegraphics[width=0.5\textwidth]{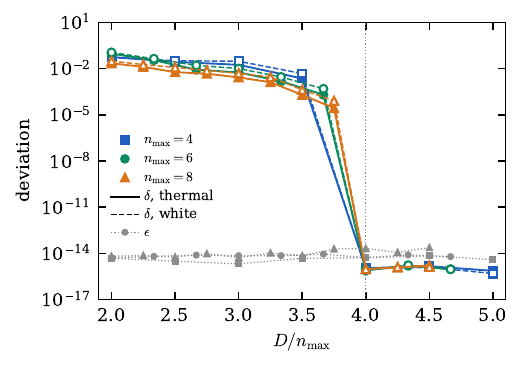}
  \caption{Deviations from $\mathrm{SO}(5)$ invariance of the interaction (coloured) and from orthonormality (grey), against $D/\nmax$, where $D$ is the degree of the quadrature grid. Coloured: $\delta=\max_R|I_4(R)-I_4|/|I_4|$, where $I_4=\sum_aw_a\phi(x_a)^4$ and $I_4(R)$ is the same sum for the field rotated by $R$, maximized over eight random $R\in\mathrm{SO}(5)$. The field has independent unit-variance Gaussian mode coefficients (white; open symbols, dashed) or variances $1/(\lambda_n+0.5)$ (free thermal; filled symbols, solid). Grey: orthonormality error of the grid, $\epsilon=\max|\sum_aw_aY_{nI}(x_a)Y_{n'I'}(x_a)-\delta_{nn'}\delta_{II'}|$. The dotted line marks $D=4\nmax$, where $\delta$ reaches double-precision roundoff at every cutoff. At $D=2\nmax$ the symmetry is broken at the percent level although orthonormality holds.}
  \label{fig:so5}
\end{figure}

Since polynomials of degree at most $4\nmax$ form an $\mathrm{SO}(5)$-invariant space, a grid with $D\ge4\nmax$ gives the same action for a field and for any rotation of it, so the regulated theory is exactly $\mathrm{SO}(5)$ invariant. Figure~\ref{fig:so5} verifies this directly and also shows the orthonormality error of the harmonics on the grid.

\subsection{Free propagator and loop sums}

By the addition theorem, Eq.~\eqref{eq:app-addition}, the free propagator at finite cutoff is
\begin{equation}
 G_{\nmax}(\vartheta;m^2)=\frac{1}{\Om}\sum_{n=0}^{\nmax}
 \frac{d_n C_n^{(3/2)}(\cos\vartheta)}
 {C_n^{(3/2)}(1)(\lambda_n+m^2)} .
\label{eq:app-prop}
\end{equation}
For positive mass, the continuum propagator is $G_\infty(\vartheta;m^2)=\Gamma(\Delta_+)\Gamma(\Delta_-)(4\pi)^{-2}{}_2F_1(\Delta_+,\Delta_-;2;\cos^2\frac\vartheta2)$, with $\Delta_\pm=\frac32\pm\sqrt{\frac94-m^2}$~\cite{Allen:1985ux}. We check Eq.~\eqref{eq:app-prop} against $G_\infty$ using the difference between two masses at noncoincident, nonantipodal points, where the sums over $n$ converge absolutely: summed through $n=400\,000$, it agrees with the continuum to $3.4\times10^{-14}$ for the tested masses and angles.

The perturbative predictions of the main text use the finite sums $S_k(\nmax)=\sum_{n=1}^{\nmax}d_n/\lambda_n^k$, not their asymptotic forms. The summand of $S_2$ is $1/(3n)+O(n^{-2})$, so
\begin{equation}
 u=\frac{9}{8\pi^2}\ln\nmax+O(1),
\end{equation}
whose coefficient, $9/8\pi^2$, is that of the flat-space one-loop $\beta$ function of the potential $\lambda\phi^4/4$, $\beta_\lambda=9\lambda^2/8\pi^2$.

\section{Monte Carlo algorithm}
\label{app:mc}

We sample the mode coefficients with hybrid Monte Carlo (HMC)~\cite{Duane:1987de}. Each trajectory draws Gaussian momenta $p_{nI}$ with variance $M_n$, integrates the equations of motion of
\begin{equation}
 \mathcal H=\frac12\sum_{nI}\frac{p_{nI}^2}{M_n}+S[c]
\end{equation}
with the leapfrog integrator for $N_{\rm md}$ steps of size $\epsilon$, and accepts the result with probability $\min(1,e^{-\Delta\mathcal H})$. The mass matrix $M_n$ implements Fourier acceleration~\cite{Batrouni:1985jn}: for $M_n=\lambda_n+m^2$, every mode of the free theory is an oscillator of unit frequency. Since the tuned bare mass is negative, $\lambda_n+m^2$ can vanish or be negative for the lowest levels, and we use $M_n=\max(\lambda_n+m^2,M_{\rm floor})$, with $M_{\rm floor}=\lambda_1+m^2$ if positive and otherwise the smallest positive $\lambda_n+m^2$. For Set~B we use instead $M_n=\max(\lambda_n,\lambda_1)$, built from the massless propagator, which remains well conditioned while the mass is tuned. The mass matrix does not change the sampled distribution.

The nominal trajectory length is $T=N_{\rm md}\epsilon=1.5$, with $\epsilon=0.25$. Since the free modes share a common frequency, a fixed $T$ would be resonant near $T\simeq\pi$, and we draw $N_{\rm md}$ uniformly within $\pm30\%$ of its nominal value for each trajectory~\cite{Mackenzie:1989us}. At strong coupling, the integrator becomes unstable on rare large-field configurations, where the local curvature $3\lambda\phi(x)^2$ is large. Each Set~B chain is therefore calibrated over $600$ trajectories, reducing $\epsilon$ by a factor $0.7$ at fixed $T$ until $\max|\Delta\mathcal H|<\ln(0.1\times600)$ and the acceptance exceeds $0.45$. This keeps $\epsilon=0.25$ at every cutoff for $\lambda\le3$, and at $\nmax=4$ and $6$ up to $\lambda=10$; at larger couplings the calibrated $\epsilon$ decreases with both $\lambda$ and $\nmax$, down to $0.005$ at $(\nmax,\lambda)=(6,60)$.

In unconstrained chains, the zero mode, the slowest direction of HMC, is updated twice more after each trajectory with an exact conditional (heat-bath) update; in the constrained ensembles of Appendix~\ref{app:force} it is held fixed. Writing $\phi(x_a)=c_0/\sqrt\Om+r_a$, with $R_k=\sum_aw_ar_a^k$, the action of $c_0$ at fixed nonzero modes is
\begin{equation}
 S(c_0|\text{rest})=\frac{\lambda}{4\Om}c_0^4
 +\left(\frac{m^2}{2}+\frac{3\lambda R_2}{2\Om}\right)c_0^2+\frac{\lambda R_3}{\sqrt\Om}c_0 .
\end{equation}
We draw a candidate by inverting the tabulated cumulative distribution of $e^{-S(c_0|\text{rest})}$ and accept it with an independence Metropolis test against the exact density, which makes the update exact.

Chains start from a free-theory Gaussian draw with variances $1/\max(\lambda_n+m^2,\min_nM_n)$ and are thermalized with the interacting action. Every production chain satisfies $0.9<\ev{e^{-\Delta\mathcal H}}<1.1$. In the free theory, the sampled mode variances agree well with $1/(\lambda_n+m^2)$, and $\ev{e^{-\Delta\mathcal H}}=0.99994$; the force agrees with finite differences of the action to better than $10^{-8}$. Figure~\ref{fig:mc} shows one chain at $(\nmax,\lambda)=(10,0.01)$, run with the Set~A parameters of Appendix~\ref{app:data}, together with three of its field configurations. An animated version is available at \url{https://caravangelo.github.io/lattice-dS/}.

\begin{figure}
  \centering
  \includegraphics[width=0.9\textwidth]{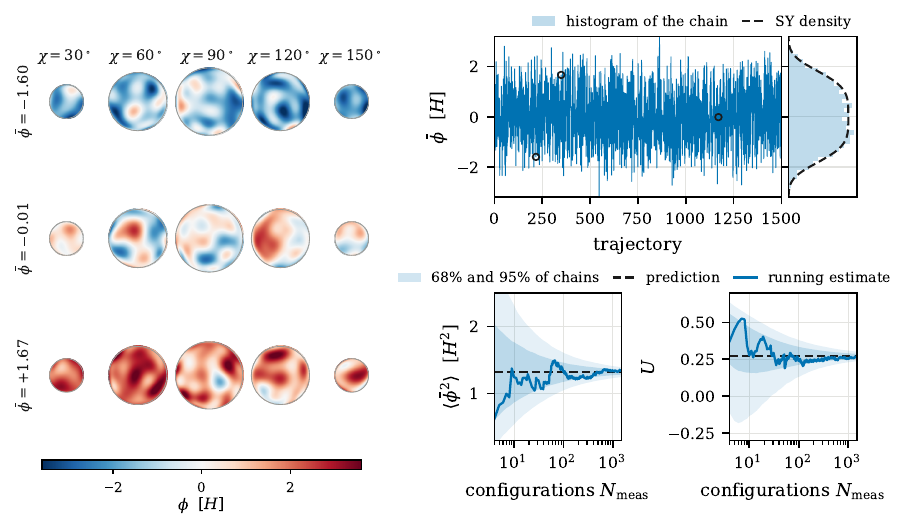}
  \caption{Monte Carlo sampling at $(\nmax,\lambda)=(10,0.01)$: one chain of $1500$ trajectories, after $2000$ thermalization trajectories. Left: three configurations of the chain, one per row, labelled by their zero mode $\phib$. Each is shown on five 2-spheres of $S^4$, at fixed $\chi$ and $\psi=\pi/2$, drawn to scale with radius $H^{-1}\sin\chi$. Top right: $\phib$ along the chain, with circles marking the three configurations, and its histogram against the SY distribution. Bottom right: estimates of $\ev{\phib^2}$ and $U$ from the first $N_{\rm meas}$ configurations of the chain, against the one-loop prediction of Eq.~\eqref{eq:prediction} and $U_\SY$; the bands contain $68\%$ and $95\%$ of independent chains of that length drawn from the SY distribution. An animated version is available at \url{https://caravangelo.github.io/lattice-dS/}.}
  \label{fig:mc}
\end{figure}

\section{The weak-coupling Set A}
\label{app:data}

In Set~A we fix the bare mass to the one-loop counterterm $m^2=-3\lambda S_1/\Om$, so that $m_R^2=0$ at one loop without fitting simulation data.

Set~A consists of $24$ points $(\lambda,\nmax)$. Each point has $18$ independent chains ($20$ for the three weakest couplings at $\nmax=10$), each with $2.5\times10^4$ measurements after $2\times10^3$ thermalization trajectories, for $438$ chains in total. At each point, we measure $\ev{\phib^2}$ and $U$ in each chain, with errors obtained from the effective number of independent measurements $N_{\rm meas}/[2\tau_{\rm int}(\phib^2)]$~\cite{Wolff:2003sm} for $\ev{\phib^2}$, and from a blocked jackknife with $100$ blocks of $250$ trajectories for $U$. We then combine the chains: the value at each point, used in all figures and fits, is the inverse-variance weighted mean of the values $v_i\pm\Delta_i$ measured in the chains $i$ at that point, with error $(\sum_i\Delta_i^{-2})^{-1/2}$. The values of $u$ in Fig.~\ref{fig:run} are fits of these points to $\ev{\phib^2}/\ev{\phib^2}_\SY-1=u\lambda/2$ at each cutoff.

The effective coupling of Eq.~\eqref{eq:lameff} at the Set~A points, shown as open symbols in Fig.~\ref{fig:triv}, is obtained with the mean-force procedure outlined in Appendix~\ref{app:setb}, at the analytic Set~A mass. It requires the mean force $\VC'(\phib)$ at $\phib=\sigma$, where $\sigma^2=\ev{\phib^2}$. We measure the force in three constrained simulations, in which the zero mode is held fixed (Appendix~\ref{app:force}), at $\phib\simeq0.8\sigma$, $\sigma$ and $1.2\sigma$, with $3\times10^4$ measurements at $\nmax=4$ and $6$, $2\times10^4$ at $\nmax=8$ and $10^4$ at $\nmax=10$. The force at $\sigma$ is interpolated between the two neighbouring chains assuming a power law, $\VC'(\phib)\propto\phib^{\,p}$ (Eq.~\eqref{eq:app-interp}). For $\lambda\le0.1$ the exponent $p$ lies between $2.99$ and $3.04$, close to the value $p=3$ of a pure quartic potential. In this range the coupling can also be read directly from the zero-mode distribution: fitting $P(\phib)$ to the SY form $\exp(-\Om\lambda_{\rm fit}\phib^4/4)$ gives $\lambda_{\rm fit}=1/(\Om\ev{\phib^4})$, which agrees with $\lamE$ within $1.3\%$. We use $\lamE$ for consistency with Set~B, where $\VC$ is no longer a pure quartic (Appendix~\ref{app:observables}).

The analytic counterterm leaves a residual mass of order $\lambda^2$, which enters the zero-mode action at relative order $\lambda^{3/2}$, the same order as the first shape corrections. This mistuning grows with $\lambda$ and becomes visible where the two sets overlap: the variance of Set~A exceeds that of Set~B, in which the mass is tuned nonperturbatively, by $0.3\%$ at $\lambda=0.2$ and by $0.9$--$3.5\%$ at $\lambda=0.5$. At $\lambda=0.5$ the mistuning also affects the shape of the distribution. The one-loop sextic term $\frac92\lambda^3S_3\phib^6$ of $\Om\VC$ predicts $U-U_\SY=0.0012$. At $\nmax=4$ we measure $U-U_\SY=0.0044(9)$ in Set~A and $0.0019(9)$ in Set~B: only the nonperturbatively tuned mass is consistent with this prediction. Therefore, at $\lambda=0.5$ and above, quantitative statements are only made using Set~B. 

\section{The strong-coupling Set B}
\label{app:setb}

At each Set~B point we determine the bare mass $m_*^2$ at which the renormalized mass $m_R^2\equiv\VC''(0)$ vanishes, and measure $\sigma$, $U$ and $\lamE$ there. The points are $\nmax=4$ at $\lambda=0.2$--$60$, $\nmax=6$ at $\lambda=0.5$--$60$, $\nmax=8$ at $\lambda=0.5$--$20$, and $\nmax=10$ at $\lambda=2$--$15$. The procedure consists of the following steps, each described in one of the subsections below: constrained simulations at a single bare mass measure $\VC'$ and $\VC''(0)$ (Appendix~\ref{app:force}); exact reweighting moves these measurements to any nearby bare mass (Appendix~\ref{app:reweight}); a fit gives $m_R^2$ as a function of the bare mass, whose zero is the massless point $m_*^2$ (Appendix~\ref{app:massless}); and $\sigma$, $U$ and $\lamE$ are evaluated at $m_*^2$ (Appendix~\ref{app:observables}).

\subsection{Constrained simulations, mean force and curvature}
\label{app:force}

Writing $\phi=\phib+\psi$, where $\psi$ contains the nonzero modes and therefore $\int\psi=0$, the action separates as
\begin{equation}
 S[\phib+\psi]=S[\psi]+\Om\Big(\tfrac12m^2\phib^2+\tfrac{\lambda}{4}\phib^4\Big)
 +\tfrac32\lambda\phib^2\!\int\!\psi^2+\lambda\phib\!\int\!\psi^3 ,
 \label{eq:app-split}
\end{equation}
since all terms proportional to $\int\psi$ vanish. At fixed $\phib$, the integrand of Eq.~\eqref{eq:VC} is then a Boltzmann weight for the nonzero modes alone. A constrained simulation samples it: we fix the zero-mode coefficient $c_0=\sqrt\Om\,\phib$ and run the HMC without the zero-mode update on the remaining $N-1$ coefficients.

The change of variables to $(\phib,\psi)$ is linear, so differentiating $\Om\VC=-\ln\int\!\mathcal D\psi\,e^{-S}$ with respect to $\phib$ gives $\VC'(\phib)=\ev{f_{\phib}}_{\phib}$, Eq.~\eqref{eq:force}, with the force
\begin{equation}
 f_{\phib}\equiv\frac1\Om\frac{\partial S}{\partial\phib}
 =m^2\phib+\lambda\phib^3+\frac{3\lambda\phib}{\Om}I_2
 +\frac\lambda\Om I_3,\qquad I_k\equiv\int_{S^4}\psi^k ,
 \label{eq:app-force}
\end{equation}
where $\ev{\cdot}_{\phib}$ is the average in the constrained simulation. In the language of constrained molecular dynamics, $\VC$ is the potential of mean force along $\phib$~\cite{Kirkwood:1935pmf,Carter:1989crc,Sprik:1998fcm}.

The massless condition requires the curvature $\VC''(0)$. The force gives the slope $\VC'$, from which the curvature could be estimated as $\VC'(\phib)/\phib$ at small $\phib$. At small $\phib$, however, the force fluctuates mainly through the term $\lambda I_3/\Om$, which does not vanish at $\phib=0$, so the statistical error of $\VC'$ is nearly independent of $\phib$ and that of $\VC'/\phib$ grows as $1/\phib$. The $\phib=0$ point cannot be reached this way. We therefore differentiate $\VC$ once more, and write Eq.~\eqref{eq:app-split} in terms of the zero-mode coefficient $c_0=\sqrt\Om\,\phib$, as a polynomial whose coefficients depend on the nonzero modes,
\begin{equation}
 S=S[\psi]+A\,c_0+\tfrac12B\,c_0^2+\frac{\lambda}{4\Om}c_0^4,\qquad
 A=\frac{\lambda}{\sqrt\Om}I_3,\qquad B=m^2+\frac{3\lambda}{\Om}I_2 ,
\end{equation}
so that $A$ and $B$ are the force on the zero mode and its stiffness at the origin. The function $W(c_0)\equiv-\ln\int\!\mathcal D\psi\,e^{-S}$ equals $\Om\VC(c_0/\sqrt\Om)$ up to a constant, so its second derivative with respect to $c_0$ is $\VC''$. Expanding $e^{-S}$ to second order in $c_0$ and averaging in the simulation at $c_0=0$, denoted $\ev{\cdot}_0$, gives $W(c_0)-W(0)=\tfrac12(\ev B_0-\ev{A^2}_0)\,c_0^2+O(c_0^4)$, and therefore
\begin{equation}
 m_R^2=\VC''(0)=\ev B_0-\ev{A^2}_0 ,
 \label{eq:app-curvature}
\end{equation}
as $\ev A_0=0$ by symmetry. This equation shows that curvature is the mean stiffness minus the variance of the force. It is the general identity $W''=\ev{\partial^2_{c_0}S}-\ev{(\partial_{c_0}S)^2}+\ev{\partial_{c_0}S}^2$ for the curvature of a free energy, the static form of the fluctuation--dissipation theorem~\cite{Kubo:1966fyg}. At strong coupling it is a small difference of two large terms, and alone it does not fix $m_R^2$ precisely enough. We therefore determine $m_R^2$ from a joint fit of this curvature and of the forces obtained from constrained simulations at nonzero $\phib$ (Appendix~\ref{app:massless}).

\subsection{Mass reweighting}
\label{app:reweight}

The force and the curvature are measured at a bare mass $m_s^2$, which is not the massless point that makes the renormalized mass vanish. Since the bare mass enters the action only through $\frac12m^2J_2$, with $J_2\equiv\int_{S^4}\phi^2$, the average of an observable $O$ at $m^2=m_s^2+\delta m^2$ follows exactly from the same configurations~\cite{Ferrenberg:1988yz,Hasenfratz:2008fg},
\begin{equation}
 \ev O_{m_s^2+\delta m^2}
 =\frac{\ev{O\,e^{-\delta m^2 J_2/2}}_{m_s^2}}
 {\ev{e^{-\delta m^2 J_2/2}}_{m_s^2}} .
 \label{eq:app-reweight}
\end{equation}
As a consequence, every stored configuration can be translated to $m^2$ if given the weight $e^{-\delta m^2 J_2/2}$. In a constrained simulation $J_2=I_2+\Om\phib^2$, and the constant term cancels, leaving the weights $e^{-\delta m^2 I_2/2}$. Observables that contain the bare mass explicitly are also shifted, $f_{\phib}\to f_{\phib}+\delta m^2\,\phib$ and $B\to B+\delta m^2$.

Equation~\eqref{eq:app-reweight} is an exact identity between ensemble averages. On a finite sample, its precision depends on how much the weights $e^{-\delta m^2 I_2/2}$ vary, which grows with $|\delta m^2|$. This is measured by the effective sample fraction $(\sum_iw_i)^2/(N_{\rm meas}\sum_iw_i^2)$~\cite{Albergo:2019eim}, which equals one for equal weights and $1/N_{\rm meas}$ when a single configuration dominates. The shift from the simulated bare mass to the massless point $m_*^2$ (Appendix~\ref{app:massless}) is at most $|\delta m^2|=0.23$, at $(\nmax,\lambda)=(6,30)$, and below $0.07$ at all other points. At these shifts the effective sample fraction is above $0.99$ in every constrained simulation, so mass reweighting is reliable throughout the analysis.

\subsection{Massless point}
\label{app:massless}

To locate the massless point, we start from a rough estimate $m_s^2$, obtained from short preliminary simulations at $\phib=0$ or from the histogram method described at the end of this subsection. At $m_s^2$ we run constrained simulations: two chains at each of $\phib=0$, $0.10$ and $0.15$, and one chain at each of $\phib\simeq0.8\sigma$, $\sigma$ and $1.2\sigma$, with $\sigma$ from unconstrained simulations at $m_s^2$. At $\lambda\ge30$ for $\nmax=4$ and $\lambda\ge15$ for $\nmax\ge6$, where the uncertainty of $m_*^2$ dominates the error of the effective coupling (Appendix~\ref{app:observables}), these are replaced by a grid of chains at $\phib=0$--$0.65$, with several replicas per value, up to $85$ chains per point. Chains have $10^5$ measurements at $\nmax=4$, $(3$--$5)\times10^4$ at $\nmax=6$ and $8$, and $10^4$ at $\nmax=10$.

Near the origin,
\begin{equation}
 \frac{\VC'(\phib)}{\phib}
 =m_R^2+\lamR^{(\phi^4)}\phib^2+g_6\phib^4,
 \label{eq:app-forcefit}
\end{equation}
where $\lamR^{(\phi^4)}$ is the quartic Taylor coefficient of $\VC$ (Appendix~\ref{app:observables}) and $g_6$ the coefficient of $\phib^6/6$. This is a parabola in $\phib^2$ whose intercept is $m_R^2$. We fit it to the mean force $\VC'=\ev{f_{\phib}}_{\phib}$ of Eq.~\eqref{eq:app-force}, divided by $\phib$, from the constrained simulations at $\phib>0$, together with the curvature of Eq.~\eqref{eq:app-curvature}, which gives the intercept directly. By default the fit includes all chains, $\phib\le1.25\sigma$. At the points with additional chains it is restricted to $\phib\le0.4$ for $\nmax=4$ and $\phib\le0.5$ for $\nmax\ge6$, which excludes the chains at larger $\phib$, where the three terms of Eq.~\eqref{eq:app-forcefit} no longer describe the data.

\begin{figure}
  \centering
  \includegraphics[width=0.82\textwidth]{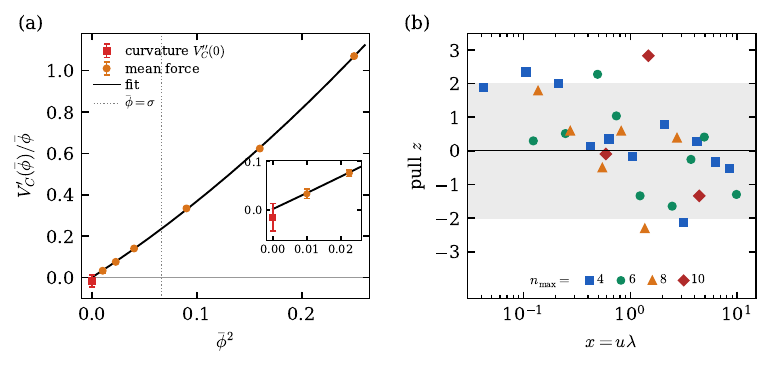}
  \caption{(a)~$\VC'/\phib$ against $\phib^2$ at $(\nmax,\lambda)=(8,20)$, after reweighting to $m_*^2$: the curvature of Eq.~\eqref{eq:app-curvature} at the origin, the mean force at $\phib>0$, and the fit of Eq.~\eqref{eq:app-forcefit} in the window $\phib\le0.5$. Chains at the same $\phib$ are combined in the figure, while the fit uses them individually. (b)~Pull of Eq.~\eqref{eq:app-pull} at the $29$ Set~B points where both estimates of the massless point are available.}
  \label{fig:massless}
\end{figure}

Reweighting every simulation to different bare masses with Eq.~\eqref{eq:app-reweight} and refitting gives $m_R^2$ as a function of the bare mass, without new simulations. Its zero is $m_*^2$ (Fig.~\ref{fig:massless}a), with a statistical error below $10^{-3}$ at $\lambda\le1$ and up to $0.02$ at the strongest couplings. To estimate the systematic error due to the choice of window, we repeat the determination with the simulations at $\phib\le0.15$ alone, fitted to $\VC'/\phib=m_R^2+b\,\phib^2$. At $(\nmax,\lambda)=(6,15)$ and $(6,20)$, which have no chains at $\phib=0.10$ and $0.15$, it instead uses Eq.~\eqref{eq:app-forcefit} in the narrower window $\phib\le0.4$. The difference between the two values of $m_*^2$ sets the systematic error of the effective coupling quoted in Appendix~\ref{app:observables}.

We validate this determination of the massless point with an independent method, which locates it without constrained simulations. To do so, we use the zero-mode distributions of unconstrained simulations at seven to eleven bare masses $m_j^2$ around $m_*^2$ (six to twelve chains of $(1.5$--$3)\times10^4$ measurements per mass), fitted jointly to the following distribution
\begin{equation}
 -\ln P_j(\phib)=\Om\!\left[\frac12m_R^2(m_j^2)\phib^2
 +\frac14\lamR^{(\phi^4)}(m_j^2)\,\phib^4+\frac16g_6\phib^6+\frac18g_8\phib^8\right]+\nu_j ,
 \label{eq:app-scanfit}
\end{equation}
with one normalization $\nu_j$ per mass and errors from a bootstrap over chains. By default, $m_R^2$ is set to be linear in the bare mass, $\lamR^{(\phi^4)}$ is independent of it, and $g_8=0$. A quadratic term in $m_R^2$, a linear mass dependence of $\lamR^{(\phi^4)}$ and $g_8$ are added only when a nested test, calibrated by a parametric bootstrap, requires them at the $1\%$ level, and a fit is used only if it passes a calibrated goodness-of-fit test. The zero of $m_R^2$ gives a second estimate $m^2_{\rm hist}$ of the massless bare mass, which we compare with $m_*^2$ through the pull
\begin{equation}
 z=\frac{m_*^2-m^2_{\rm hist}}{\sqrt{\Delta_*^2+\Delta_{\rm hist}^2}} ,
 \label{eq:app-pull}
\end{equation}
where $\Delta_*$ and $\Delta_{\rm hist}$ are the statistical errors of the two estimates. If both are unbiased and their errors correct, $z$ follows a standard normal distribution, with zero mean and unit width. The histogram estimate is available at $29$ points, $\lambda=0.2$--$40$, where the two masses differ by at most $0.045$. The pulls, shown in Fig.~\ref{fig:massless}(b), have mean $0.23\pm0.25$, so the two independent determinations agree well. Their scatter does not seem to be random: below $x\simeq1$ the histogram estimate lies systematically below $m_*^2$, and six points lie outside $|z|\le2$, where a standard normal distribution would place about one. We attribute this to the histogram fit, whose $\phib^2$ coefficient is poorly constrained at weak coupling, where the quartic term dominates $P(\phib)$. The mean-force result shows no such feature: its deviation from the one-loop bare mass varies smoothly with $\lambda$ at every cutoff, as $\lambda^2$ at weak coupling.

\subsection{Effective coupling and Binder cumulant}
\label{app:observables}

At $m_*^2$ we measure the width $\sigma=\sqrt{\ev{\phib^2}}$, the Binder cumulant $U$, and the effective coupling $\lamE$ of Eq.~\eqref{eq:lameff}, plotted in the main text as $y=u\lamE$. The first two come from unconstrained simulations at a single bare mass close to $m_*^2$, reweighted to $m_*^2$.
The effective coupling requires the mean force at $\phib=\sigma$. The constrained simulations at $\phib\simeq0.8\sigma$, $\sigma$ and $1.2\sigma$ were placed using a preliminary estimate of $\sigma$, so none of them lies exactly at the final value. Instead of running a new constrained simulation at the final value of $\sigma$, we take the two that bracket it, $\phib_-<\sigma<\phib_+$, from the constrained set. We reweight them to $m_*^2$, and interpolate between their forces $F_\pm$ with a local power law
\begin{equation}
 p=\frac{\ln(F_+/F_-)}{\ln(\phib_+/\phib_-)},\qquad
 F_\sigma=F_-\left(\frac{\sigma}{\phib_-}\right)^p,\qquad
 y=u\,\frac{F_\sigma}{\sigma^3},
 \label{eq:app-interp}
\end{equation}
which is exact for a pure quartic potential ($p=3$).

We use $\lamE$ rather than the quartic Taylor coefficient $\lamR^{(\phi^4)}=\VC^{(4)}(0)/6$, which defines $\lamR$ in the one-loop analysis of the main text. The two differ by the higher-order terms of $\VC$ evaluated at $\sigma$,
\begin{equation}
\lamE=\lamR^{(\phi^4)}+g_6\sigma^2+\cdots ,
 \label{eq:app-couplings}
\end{equation}
as $\VC=\lamR^{(\phi^4)}\phib^4/4+g_6\phib^6/6+\cdots$ at the massless point. They agree within a few per cent for $x\lesssim1$. At strong coupling, however, the higher terms are not small at $\phib\simeq\sigma$, and extracting the Taylor coefficient requires extrapolating a polynomial fit to $\phib=0$, whose result is sensitive to the number of higher-order terms included in the fit. The effective coupling, instead, is simpler to extract and has a direct physical meaning: it is the quartic coupling felt by the zero mode at its typical field $\sigma$. We restrict Set~B to $\lambda\le60$. At larger couplings the massless point becomes harder to determine, and we leave this regime to future work.

Statistical errors come from a joint block bootstrap. Every chain is resampled in blocks of $500$ measurements, and the whole analysis, including the determination of $m_*^2$, is repeated in each replica. The error of $y$ therefore includes the uncertainty of $m_*^2$, which dominates it at strong coupling: a shift $\delta m^2$ of the bare mass changes the force at $\sigma$ by $\delta m^2\sigma$, and hence $y$ by $u\,\delta m^2/\sigma^2\sim4\,\delta m^2$. As a systematic error, we take half the change of $y$ when $m_*^2$ is replaced by its second determination, which uses a different fit window (Appendix~\ref{app:massless}). Table~\ref{tab:setb2} lists the points with $x>3$. The constant fit $y_\infty=0.89\pm0.01$ of the main text uses the ten points with $x>4$, at all four cutoffs, with the two errors of the table added in quadrature, giving $\chi^2/{\rm dof}=9.6/9$. Moving the cut between $x>3.5$ and $x>6$ changes $y_\infty$ by at most $0.005$.

\begin{table}
\centering
\caption{Set~B points with $x>3$. The first error on $y$ and the error on $U$ come from the joint bootstrap, which includes the statistical uncertainty of $m_*^2$; the second error on $y$ is the systematic error from the choice of fit window (Appendix~\ref{app:massless}).}
\label{tab:setb2}
\begin{tabular}{ccccccc}
\hline\hline
$\nmax$ & $\lambda$ & $x$ & $x/(1+x)$ & $y$ & $U$ & $\sigma$ \\
\hline
4 & 15 & 3.2 & 0.759 & $0.794(22)(3)$ & $0.2908(24)$ & 0.250 \\
4 & 20 & 4.2 & 0.808 & $0.849(33)(4)$ & $0.2973(25)$ & 0.245 \\
4 & 30 & 6.3 & 0.863 & $0.894(12)(11)$ & $0.2987(32)$ & 0.239 \\
4 & 40 & 8.4 & 0.894 & $0.882(16)(0)$ & $0.2996(22)$ & 0.236 \\
4 & 60 & 12.6 & 0.927 & $0.903(48)(72)$ & $0.3097(31)$ & 0.235 \\
6 & 15 & 3.7 & 0.787 & $0.856(25)(11)$ & $0.2923(29)$ & 0.260 \\
6 & 20 & 4.9 & 0.831 & $0.902(33)(6)$ & $0.2941(32)$ & 0.254 \\
6 & 30 & 7.4 & 0.881 & $0.943(40)(63)$ & $0.2913(54)$ & 0.248 \\
6 & 40 & 9.9 & 0.908 & $0.943(48)(22)$ & $0.2943(37)$ & 0.246 \\
6 & 60 & 14.8 & 0.937 & $0.928(61)(61)$ & $0.3025(53)$ & 0.246 \\
8 & 20 & 5.5 & 0.845 & $0.995(38)(19)$ & $0.2863(75)$ & 0.257 \\
10 & 15 & 4.4 & 0.815 & $0.855(70)(52)$ & $0.2907(46)$ & 0.269 \\
\hline\hline
\end{tabular}
\end{table}

\FloatBarrier

\end{document}